\documentclass[aps,reprint,groupedaddress,showpacs,longbibliography,nofootinbib]{revtex4-2}
\usepackage{amssymb}
\usepackage{amsmath}
\usepackage[breaklinks,colorlinks,citecolor=green,urlcolor=blue,linkcolor=red]{hyperref}
\usepackage{graphicx}
\usepackage{epstopdf}
\usepackage{diagbox}
\usepackage{float}
\usepackage{subfigure}
\usepackage{multirow}
\usepackage{CJK}
\usepackage{tikz}
\usepackage{enumerate}

\begin{document}


\title{\boldmath How to measure the transverse polarization of the produced hadrons in a symmetric collider?}


\author{Zhen-Hua Zhang}
\email{zhangzh@usc.edu.cn}
\affiliation{School of Nuclear Science and Technology, University of South China, Hengyang, 421001, Hunan, China}


\date{\today}
\begin{abstract}
In this paper, some subtleties in the measurement of the transverse polarization of the produced hadrons on symmetric colliders---such as the Large Hadron Collider when conducting the $pp$ collisions---are revealed.
It can be proved that the transverse polarization of the produced particles with opposite pseudorapidity takes exactly opposite values if the normal vector of the production plane is defined in a convention-dependent way, regardless of whether parity is conserved or not in the production process. 
The analysis shows that, due to the symmetry of the initial state, the $\Lambda_b$ transverse polarization measured by the CMS collaboration in Phys. Rev. D 97, 072010 (2018) 
should be exactly equal to zero. 
A modified measurement of the $\Lambda_b$ polarization for CMS and ATLAS is proposed, the result of which can be compared to the LHCb measurement.
\end{abstract}
\maketitle

Polarized hadrons have important applications in experimental research.
The transversely polarized hyperon, for example, was used to test the parity violation in its decay processes \cite{Lee:1957he,Eisler:1957ih,Crawford:1957zzb}.
The polarization of hadrons is an important probe of properties of QGP in heavy-ion collisions \cite{STAR:2017ckg}.
A number of New Physics models are sensitive to the polarization of the particles involved \cite{Krohn:2011tw}.
Therefore, the production of particles with large polarization, as well as the measurements of the corresponding polarization, are experimental tasks of great importance.

It is well known that due to the parity symmetry of the production process, strongly and electromagnetically produced particles can only have none-zero transverse polarization for unpolarized initial state \cite{Lee:1957he}.
Transverse polarization of different baryons on several colliders, such as $\Lambda$, $\Lambda_c$, and $\Lambda_b$, was measured by different collaborations \cite{Bunce:1976yb,Belle:2018ttu,BESIII:2019odb,LHCb:2013hzx,CMS:2018wjk}.
Besides, the longitudinal polarization of the top quark in pair \cite{CMS:2013roq,ATLAS:2016bac} and single production \cite{CMS:2015cyp,ATLAS:2017ygi} was also measured by CMS and ATLAS collaborations.
In particular, the transverse polarization of $\Lambda_b$ was measured by LHCb and CMS collaborations successively through the decay channel $\Lambda_b\to J/\psi \Lambda$.
According to the measurements, the transverse polarization of the produced $\Lambda_b$ is measured to be
\begin{equation}\label{eq:CMS}
  P_{\Lambda_b}^{\text{CMS}}=0.00\pm0.06\pm0.06
\end{equation}
by CMS \cite{CMS:2018wjk}, and was claimed to be consistent with the LHCb measurement \cite{LHCb:2013hzx}:
\begin{equation}\label{eq:LHCb}
  P_{\Lambda_b}^{\text{LHCb}}=0.06\pm0.07\pm0.02.
\end{equation}
Both of the two aforementioned measurements of the $\Lambda_b$ transverse polarizations are adopted by the Heavy Flavour Averaging Group to obtain an average value 
\begin{equation}
  P_{\Lambda_b}^{\text{HFLAV}}=0.03\pm0.06.
\end{equation}
While the theoretical prediction for the transverse polarization of $\Lambda_b$ in hadron colliders is at the 10\% level \cite{Dharmaratna:1996xd,Hiller:2007ur}.
We want to point out, however, that the measurement of CMS in Eq. (\ref{eq:CMS}) should not be regarded as a cross check to the LHCb result in Eq. (\ref{eq:LHCb}).

CMS is a detector built on the Large Hadron Collider (LHC), which conducts mainly unpolarized symmetric $pp$ collisions. 
LHC is a symmetric collider in the sense that the initial state $pp$ is unchanged under the parity transformation centered on the collision point.
Moreover, the initial state is also unchanged under a rotation of an angle $\pi$ around any axis intersecting with the beam line at the collision point that is perpendicular to the beam line. 
This has important constraints to the polarization measurement, which will be shown in what follows.

The transverse polarization of the produced hadrons is defined along the normal vector to the decay plane, which is usually chosen by convention. 
For example, the transverse polarization of $\Lambda_b$ produced on CMS is defined along the normal vector
\begin{equation}\label{eq:n}
  \hat{n}=\frac{\vec{p}_{\text{beam}}\times \vec{p}_{\Lambda_b}}{|\vec{p}_{\text{beam}}\times \vec{p}_{\Lambda_b}|}
\end{equation}
to the production plane, 
where $p_{\Lambda_b}$ is the momentum of $\Lambda_b$ in the laboratory frame,  $\vec{p}_{\text{beam}}$ is chosen as the one which transfers counterclockwise in the beam pipe \cite{CMS:2008xjf}.
It can be shown that for the same transverse momentum $p_T$,
the differential transverse polarization of $\Lambda_b$ is an odd function of the pseudorapidity $\eta$  \cite{ATLAS:2014swk}:
\begin{equation}\label{eq:ATLAS}
  P_{\hat{n}}(\eta,p_T)=-P_{\hat{n}}(-\eta,p_T),
\end{equation}
where $P_{\hat{n}}\equiv \hat{n}\cdot \vec{P}_{\Lambda_b}$ is the polarization of $\Lambda_b$ along $\hat{n}$, with $\vec{P}_{\Lambda_b}$ being the polarization vector of $\Lambda_b$, 
and the pseudorapidity $\eta$ is defined as $\eta\equiv -\ln \left[\tan \left(\frac{\theta}{2}\right)\right]$, with $\theta$ being the polar angle of $\vec{p}_{\Lambda_b}$ with respect to $\vec{p}_{\text{beam}}$.
Since the transverse momentum $p_T$ is irrelevant to the main points we discuss here, we will simply ignore the $p_T$-dependence of the transverse polarization in what follows.

It should be pointed out that Eq. (\ref{eq:ATLAS}) is always true for a symmetric collider regardless of whether parity symmetry is respected  the production process.
Proof of this statement is presented explicitly in the Appendix.
Consequently, the integrated transverse polarization on a symmetric interval of $\eta$ should be equal exactly to zero: 
\begin{equation}\label{eq:CMS}
  \langle P_{\hat{n}}\rangle^{\eta\in (-a,a)} \equiv\frac{\int_{-a}^{+a} P_{\hat{n}}(\eta) N(\eta) d\eta}{\int_{-a}^{+a} N(\eta)d\eta}=0,
\end{equation}
where $N(\eta)$ is the event yields in the interval from $\eta$ to $\eta+d\eta$ and is an even function of $\eta$.
One then concludes that the transverse polarization of $\Lambda_b$ measured by CMS in Ref. \cite{CMS:2018wjk} should equal exactly to zero, since CMS is a detector covering a symmetric region of pseudorapidity \cite{CMS:2008xjf}, and the contributions with opposite pseudorapidities cancel exactly.

The cancellation problem can be resolved if the chosen interval of $\eta$ is not symmetric around 0.
One typical example is the transverse polarization of the forward region of the pseudorapidity $\eta$ \footnote{Strictly speaking, there is no distinction between forward and backward for a symmetric collider. The ``forward region'' here refers to $\eta>0$, where $\eta$, as is indicated in the main text, is defined according to the conventionally chosen beam direction $\vec{p}_{\text{beam}}$.}:
\begin{equation}\label{eq:forward}
  P_{\Lambda_b}^{\text{forward}}\equiv\frac{\int_{0}^{+\infty} P_{\hat{n}}(\eta) N(\eta) d\eta}{\int_{0}^{+\infty} N(\eta)d\eta}.
\end{equation}
In reality, a detector cannot cover the whole $\eta>0$ region.
For example, the LHCb covers a pseudorapidity range of a very forward region $2<\eta<5$ \cite{AbellanBeteta:2020amj}, while
the Inner Tracker system of CMS and ATLAS cover a pseudorapidity range $-2.5<\eta<2.5$ \cite{CMS:2008xjf,ATLAS:2008xda}.
Hence the LHCb is able to measure the $\Lambda_b$ transverse polarization in the very forward region:
\begin{equation}\label{eq:LHCb}
  P_{\Lambda_b}^{\text{LHCb}}= P_{\Lambda_b}^{\text{very forward}}=\frac{\int_{2}^{5} P_{\hat{n}}(\eta) N(\eta) d\eta}{\int_{2}^{5} N(\eta)d\eta},
\end{equation}
while the CMS and ATLAS is capable to measure  the $\Lambda_b$ transverse polarization in the centered pseudorapidity region:
\begin{equation}\label{eq:CMSATLAS}
P_{\Lambda_b}^{\text{central region}}=\frac{\int_{0}^{2.5} P_{\hat{n}}(\eta) N(\eta) d\eta}{\int_{0}^{2.5} N(\eta)d\eta}.
\end{equation}
One can see that the measurements of the transverse polarization of $\Lambda_b$ on CMS and ATLAS are mainly complementary to that on LHCb.
To obtain a whole picture of the transverse polarization of $\Lambda_b$ on $pp$ collisions of LHC, both the measurements of CMS/ATLAS and LHCb are essential.

The event yields in the backward region ($\eta<0$) can also be included to the measurement of $P_{\Lambda_b}^{\text{forward}}$, thus doubling the statistics.
To do so, one just need to introduce an extra signature $\text{sign}(\eta)$, so that $P_{\Lambda_b}^{\text{forward}}$ can be expressed as
\begin{equation}
  P_{\Lambda_b}^{\text{forward}}=\frac{\int_{-\infty}^{+\infty} \text{sign}(\eta) P_{\hat{n}}(\eta) N(\eta) d\eta}{\int_{-\infty}^{+\infty} N(\eta)d\eta}.
\end{equation}
One see immediately that 
the transverse polarization of the forward region can be measured on symmetric detectors such as CMS and ATLAS with the data of the full pseudorapidity region included. 

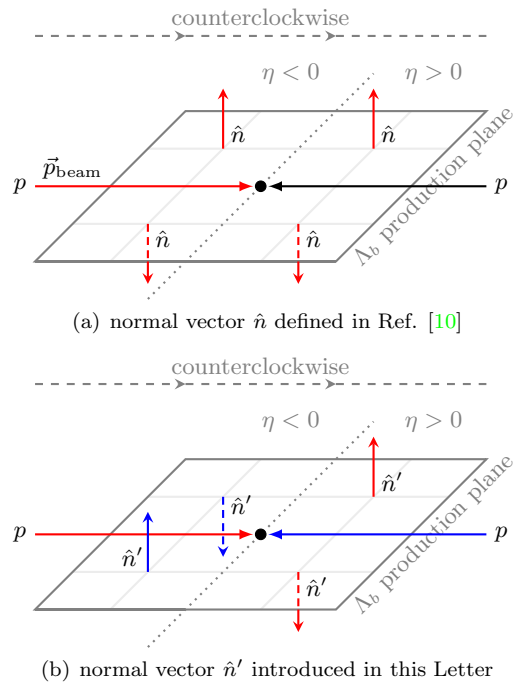
\begin{figure}[b]\centering
\subfigure[normal vector $\hat{n}$ defined in Ref. \cite{CMS:2018wjk}]{
\begin{tikzpicture}

 \draw[ thick, gray!15] (-2.5,-0.5) -- (1.5,-0.5);
 \draw[ thick, gray!15] (-1.5,0.5) -- (2.5,0.5);
 \draw[ thick,gray!15] (-2,-1) -- (0,1); 
  \draw[ thick,gray!15] (0,-1) -- (2,1);

\draw[gray,thick] (-3,-1)--(1,-1)--(3,1)--(-1,1)--(-3,-1)--(1,-1); \node [gray, rotate around={45:(0,0)}] at(2.3,0) {$\Lambda_b$ production plane};
\draw[dotted, gray, thick] (-1.5,-1.5) -- (1.5,1.5);
\node[gray] at(2.3,1.5) {$\eta>0$};\node[gray] at(0.4,1.5) {$\eta<0$};

\draw[-stealth,dashed,gray,thick](-3,2)--(-1,2);\draw[-stealth,dashed,gray,thick](-1,2)--(1,2);\draw[dashed,gray,thick](1,2)--(3,2); \node[gray]at (0,2.25) {counterclockwise};

\draw[-latex, black, thick,red] (-3,0) -- (-0.1,0);\node[] at(-3.2,0) {$p$}; \node[]at (-2.5,0.25){$\vec{p}_{\text{beam}}$};
\draw[-latex, black, thick] (3,0) -- (0.1,0); \node[] at(3.2,0){$p$};
\filldraw[black] (0,0) circle (2pt); 

\draw[-stealth, black, thick,red] (-1.5,-1) -- (-1.5,-1.3);\draw[densely dashed, black, thick,red] (-1.5,-0.5) -- (-1.5,-1);\node[] at(-1.3,-0.7) {$\hat{n}$};
\draw[-stealth, black, thick,red] (0.5,-1) -- (0.5,-1.3);\draw[densely dashed, black, thick,red] (0.5,-0.5) -- (0.5,-1);\node[] at(0.7,-0.7) {$\hat{n}$};
\draw[-stealth, black, thick,red] (1.5,0.5) -- (1.5,1.3);\node[] at(1.7,0.7) {$\hat{n}$};
\draw[-stealth, black, thick,red] (-0.5,0.5) -- (-0.5,1.3);\node[] at(-0.3,0.7) {$\hat{n}$};

\end{tikzpicture}
}
\subfigure[normal vector $\hat{n}'$ introduced in this Letter]{
\begin{tikzpicture}

 \draw[ thick, gray!15] (-2.5,-0.5) -- (1.5,-0.5);
 \draw[ thick, gray!15] (-1.5,0.5) -- (2.5,0.5);
  \draw[ thick,gray!15] (-2,-1) -- (0,1); 
  \draw[ thick,gray!15] (0,-1) -- (2,1);

\draw[gray,thick] (-3,-1)--(1,-1)--(3,1)--(-1,1)--(-3,-1)--(-1,-1); \node [gray,rotate around={45:(0,0)}] at(2.3,0) {$\Lambda_b$ production plane};
\draw[dotted, gray, thick] (-1.5,-1.5) -- (1.5,1.5);
\node[gray] at(2.3,1.5) {$\eta>0$};\node[gray] at(0.4,1.5) {$\eta<0$};

\draw[-stealth,dashed,gray,thick](-3,2)--(-1,2);\draw[-stealth,dashed,gray,thick](-1,2)--(1,2);\draw[dashed,gray,thick](1,2)--(3,2); \node[gray]at (0,2.25) {counterclockwise};

\draw[-latex, black, thick,red] (-3,0) -- (-0.1,0);\node[] at(-3.2,0) {$p$};
\draw[-latex, black, thick,blue] (3,0) -- (0.1,0); \node[] at(3.2,0){$p$};
\filldraw[black] (0,0) circle (2pt);

\draw[-stealth, black, thick,red] (0.5,-1) -- (0.5,-1.3);\draw[densely dashed, black, thick,red] (0.5,-0.5) -- (0.5,-1);\node[] at(0.75,-0.7) {$\hat{n}'$};
\draw[-stealth, black, thick,red] (1.5,0.5) -- (1.5,1.3);\node[] at(1.75,0.7) {$\hat{n}'$};

\draw[-stealth, black, thick,blue] (-1.5,-0.5) -- (-1.5,0.3);\node[] at(-1.7,-0.3) {$\hat{n}'$};
\draw[-stealth, densely dashed, black, thick,blue] (-0.5,0.5) -- (-0.5,-0.3);\node[] at(-0.25,0.4) {$\hat{n}'$};

\end{tikzpicture}
  }
\caption{Illustration of two different choices of the normal vectors. After produced, $\Lambda_b$ can fly into one of the four quadrants on the production plane. These two figures show two different choices of the normal vectors in the four quadrants. In figure (a), the normal vectors $\hat{n}$ (see Eq. (\ref{eq:n})) are always defined according to the beam that flies counterclockwise in the beam pipe, as is indicated by red arrows. While in figure (b), the normal vectors $\hat{n}^\prime$ that are indicated by red (blue) , are defined in Eq. (\ref{eq:nprime}) according to the beam in red (blue).
Notice that the pseudorapidity in both cases is always defined conventionally according to direction of $\vec{p}_{\text{beam}}$ .
}\label{fig:NV}
\end{figure}

The introduction of $\text{sign}(\eta)$ in $P_{\Lambda_b}^{\text{forward}}$ is equivalent to chose the normal vector in a {\it convention-independent} way: 
for a $\Lambda_b$ that is produced, the beam direction is chosen aligned with the momentum of $\Lambda_b$, $\vec{p}_{\Lambda_b}\cdot \vec{p}_{\text{beam}}^{~\prime}>0$, and the normal vector, which will be denoted as $\vec{n}'$, is defined as
\begin{equation}\label{eq:nprime}
  \vec{n}'\equiv \frac{\vec{p}_{\text{beam}}^{~\prime}\times \vec{p}_{\Lambda_b}}{|\vec{p}_{\text{beam}}^{~\prime}\times \vec{p}_{\Lambda_b}|}.
\end{equation}
The difference between the normal vectors $\hat{n}'$ and $\hat{n}$ is illustrated in FIG. \ref{fig:NV}. 
One can easily see that the transverse polarization of $\Lambda_b$ along $\vec{n}'$, which is defined as $P_{\hat{n}'}\equiv \hat{n}'\cdot \vec{P}_{\Lambda_b}$, is now an even function of $\eta$:
\begin{equation}\label{eq:Pnprime}
  P_{\hat{n}'}(\eta)=P_{\hat{n}'}(-\eta).
\end{equation}
The forward transverse polarization can then be re-expressed as
\begin{equation}
  P_{\Lambda_b}^{\text{forward}}=\frac{\int_{-\infty}^{+\infty}  P_{\hat{n}'}(\eta) N(\eta) d\eta}{\int_{-\infty}^{+\infty} N(\eta)d\eta}.
\end{equation}

One can also use the $\overline{\Lambda}_b$ data to further double the statistics, if CP invariance of the production process is assumed.
It should be noted that the normal vector $\overline{\vec{n}'}$ for the production plane of $\overline{\Lambda}_b$ is defined as 
\begin{equation}
  \overline{\vec{n}'}\equiv \frac{\overline{\vec{p}_{\text{beam}}^{~\prime}}\times \vec{p}_{\overline{\Lambda_b}}}{|\overline{\vec{p}_{\text{beam}}^{~\prime}}\times \vec{p}_{\overline{\Lambda_b}}|},
\end{equation}
with the beam direction chosen according to $\overline{\vec{p}_{\text{beam}}^{~\prime}}\cdot \vec{p}_{\overline{\Lambda_b}}>0$.
If the statistics is large enough, one can also measure the transverse polarization in even smaller $\eta$ intervals, so that $\Lambda_b$ with large polarization can be selected for future use.

In principle, the transverse polarization of the top quark in the forward region on the $pp$ colliders are also  worth measuring.
However, theoretical prediction of the transverse polarization of the top quark in the forward region is at the level of a few thousandths \cite{Bernreuther:2015yna}, and is out of the reach of the current experimental precision. 

To sum up, we have revealed some subtleties for the measurement of transverse polarization of the produced hadrons in symmetric colliers such as LHC. 
According to the analysis, the measurement of the $\Lambda_b$ polarization by CMS collaboration in Ref. \cite{CMS:2018wjk} should be exactly equal to zero, as is indicated by Eq. (\ref{eq:ATLAS}).
Moreover, we strongly suggest our experimental colleagues of CMS and ATLAS to perform the measurement of the $\Lambda_b$ transverse polarization in the forward region of the pseudorapidity. 
The results can be compared with the LHCb's measurement in Ref. \cite{LHCb:2013hzx}.
An improved measurement of the $\Lambda_b$ polarization by LHCb is also expected. 

\begin{acknowledgments}
I thank Hai-Bo Li, Zong-Guo Si, Chengping Shen, Longke Li and Hsiang-nan Li for valuable discussions.
I also thank the Referee for the constructive comments and suggestions, which helped me a lot on the improvement of the manuscript.
This work was supported by National Natural Science Foundation of China under Grants No. 12192261, Natural Science Foundation of
Hunan Province under Grants No. 2022JJ30483, and Scientific Research Fund of Hunan Provincial Education Department under Grants No. 22A0319.
\end{acknowledgments}

\appendix
\section{Proof of Eq. (\ref{eq:ATLAS})} 

In this Appendix, we present the proof of Eq. (\ref{eq:ATLAS}) for a symmetric collider.
Although the proof is quite straightforward, a detailed presence of the proof and a comparison with asymmetric colliders will be instructive.
As will be seen, one only needs to adopt the rotational invariance of the production process, which is always true regardless of the production process being through weak or strong interactions.
The proof is as follows.

Up to an irrelevant overall factor, the transverse polarization of $\Lambda_b$ for a given pseudorapidity $\eta$ is defined as
\begin{equation}
  P_{\hat{n}}(\eta,p_T) \!\equiv \! \sum_{\lambda,X} \lambda \left|\left\langle \Lambda_b(\vec{p}_{\Lambda_b},\lambda) X |\mathcal{T}|p(\vec{p}_{\text{beam}})p(-\vec{p}_{\text{beam}})\right\rangle\right|^2,
\end{equation}
where $\mathcal{T}$ is the transition matrix, $\lambda$ is the spin quantization of $\Lambda_b$ along the $\hat{n}$ direction, $X$ represents the collection of all the particles in the final state except $\Lambda_b$, and the summation over all the quantum numbers and the integration over all the momenta of these particles in $X$ is understood, and is shorthanded by $\sum_X$, the polarization of the proton in the initial state is omitted since the initial state is assumed to be unpolarized.
Note that although $\lambda$ remains unchanged under parity transformation, it  will transform into $-\lambda$ under the rotation of angle $\pi$ around the axis (which will be denoted as the $z$-axis hereafter) perpendicular to the production plane and intersecting with the production plane on the collision point. 
The transverse polarization for pseudorapidity with opposite value $-\eta$ can then be expressed as
\begin{equation}
  P_{\hat{n}}( \! -\eta,p_T) \!\equiv\!\! \sum_{\lambda,X} \! \lambda \! \left|\left\langle \Lambda_b ( \!-\vec{p}_{\Lambda_b},\lambda ) X |\mathcal{T}|p(\vec{p}_{\text{beam}}\!)p(-\vec{p}_{\text{beam}}\!)\right\rangle\right|^2. 
\end{equation}
One has
\begin{eqnarray}
  &&P_{\hat{n}}(-\eta,p_T) \nonumber\\&=&\sum_{\lambda,X} \lambda \left|\left\langle \Lambda_b(-\vec{p}_{\Lambda_b},\lambda) X |\mathbb{R}^{\dagger}\mathbb{R}\mathcal{T}|p(\vec{p}_{\text{beam}})p(-\vec{p}_{\text{beam}})\right\rangle\right|^2\nonumber\\
    &=&\sum_{\lambda,X} \lambda \left|\left\langle \Lambda_b(\vec{p}_{\Lambda_b},-\lambda) X^{\mathbb{R}} |\mathbb{R}\mathcal{T}|p(\vec{p}_{\text{beam}})p(-\vec{p}_{\text{beam}})\right\rangle\right|^2\nonumber\\
  &=&\sum_{\lambda} \lambda \sum_{X}\left|\left\langle \Lambda_b(\vec{p}_{\Lambda_b},-\lambda) X^{\mathbb{R}} |\mathcal{T}\mathbb{R}|p(\vec{p}_{\text{beam}})p(-\vec{p}_{\text{beam}})\right\rangle\right|^2\nonumber\\
  &=&\sum_{\lambda,X} \lambda \left|\left\langle \Lambda_b(\vec{p}_{\Lambda_b},-\lambda) X |\mathcal{T}|p(-\vec{p}_{\text{beam}})p(\vec{p}_{\text{beam}})\right\rangle\right|^2\nonumber\\
    &=&-\sum_{\lambda,X} \lambda \left|\left\langle \Lambda_b(\vec{p}_{\Lambda_b},\lambda) X |\mathcal{T}|p(\vec{p}_{\text{beam}})p(-\vec{p}_{\text{beam}})\right\rangle\right|^2\nonumber\\ &=&-P_{\hat{n}}(\eta,p_T), 
\end{eqnarray}
where in the first step the identity $\mathbb{R}^{\dagger}\mathbb{R}=1$ is inserted, with $\mathbb{R}$ being the abbreviation of $\mathbb{R}_z(\pi)$ which represents the unitary operator corresponding to the rotation of angle $\pi$ around the $z$-axis, in the second step note that $\mathbb{R}$ will change the signatures of both $\vec{p}_{\Lambda_b}$ and $\lambda$, in the third step the rotational invariance of $\mathcal{T}$ is adopted, in the fourth step we have used the fact that both the density matrix of the initial state and the $X$ part of the final state is invariant under $\mathbb{R}_{z}(\pi)$: $[\hat{\rho},\mathbb{R}]=0$, with $\hat{\rho}$ can be either $|p(\vec{p}_{\text{beam}})p(-\vec{p}_{\text{beam}})\rangle\langle p(\vec{p}_{\text{beam}})p(-\vec{p}_{\text{beam}})|$ or $\sum_X|X\rangle\langle X|$, in the fifth step $\lambda$ is replaced by $-\lambda$.

It should be pointed out that the above proof also apply to the case when $\Lambda_b$ is produced secondarily \footnote{Recent experimental study showed that the fraction for the secondary production of $\Lambda_b$ is quite small \cite{LHCb:2023tma}.}.
In this situation, the beam line may no longer lie in the decay plane. Instead, it may parallel to the production plane.

The above proof indicates that Eq. (\ref{eq:ATLAS}) always holds, regardless that $\Lambda_b$ is produced though strong or weak interactions. 
In particular, parity conservation is not required for the production process.
Eq. (\ref{eq:Pnprime}) can also be proven in a similar way, one just needs to keep in mind that the spin quantization along $\hat{n}^\prime$ is unchanged under $\mathbb{R}_z(\pi)$.

It is interesting to compare the above result with the transverse polarization of the hyperons in $e^+e^-\to\Lambda\bar{\Lambda}$ pair-production on an anti-symmetric detectors such as BESIII.
Although the transverse polarization of the hyperons are also odd functions of the pseudorapidity, as has been observed by BESIII \cite{BESIII:2022qax,BESIII:2023drj}, but this is true only when the hyperons are produced through pair-productions, as is indicated theoretically in Ref. \cite{Dubnickova:1992ii}.

\bibliography{CMSpolLbbib}
\end{document}